\documentclass[11pt, a4paper]{article}
\pdfoutput=1

\usepackage{graphicx}
\usepackage{epsfig,amsmath,amsthm}
 \usepackage{epstopdf}

\newcommand{\be}{\begin{equation}}
\newcommand{\ee}{\end{equation}}
\newcommand{\bea}{\begin{eqnarray}}
\newcommand{\eea}{\end{eqnarray}}
\def\bse{\begin{subequations}}
\def\ese{\end{subequations}}

\def\IZ{\relax\ifmmode\hbox{Z\kern-.4em Z}\else{Z\kern-.4em Z}\fi}

\def\presub{\vspace{.5cm} \noindent}

\def\bi{\begin{itemize}} \def\ei{\end{itemize}}

\def\({\left(} \def\){\right)}
\def\[{\left[} \def\]{\right]}
\def\<{\left<} \def\>{\right>}

\title{The K--Pg event as a key to bat evolution}

\author{Barak Kol  \\
{\it Racah Institute of Physics, Hebrew University}\\ {\it  Jerusalem 91904, Israel} \\
{\tt barak.kol@mail.huji.ac.il}
}

\date{\vspace{-5ex}}

\begin{document}
\maketitle

\begin{abstract}
Bats are unique mammals. This note discusses some questions regarding bat evolution including why they are nocturnal and why they can echolocate. It is hypothesized that echolocation was necessary for bats to survive the period of limited visibility that followed the Cretaceous--Paleogene (K--Pg) extinction event. 
\end{abstract}

\noindent {\bf Introduction}. Bats are amazing and unique animals: they are mammals which can fly and echolocate. There are over 1200 bat species \cite{wiki:bat,Swartz}, out of some 5500 mammal species \cite{wiki:mammal} making them the second largest order of mammals (after rodents). Some 70\% of bat species feed on insects, namely are insectivores. 

From an evolutionary viewpoint it is natural to wonder about the origin of bats and ask: when 
 and under what conditions did bats evolve? Further questions are more bat specific.  What is the relation between the evolution of flight and that of echolocation? Why are bats nocturnal?  Why are bats so relatively diverse among mammals?  In short, we are asking: what is the reason for bats? By this we mean the conditions which made its defining characteristics necessary for survival.

Already in the \emph{Origin of Species} Darwin \cite{Darwin} recognized the problem presented by the sudden appearance of `completely developed' bats. Echolocation in bats, discovered in 1958 \cite{echolocation}, aggravates the problem, by requiring an explanation for the evolution of two related traits rather than one, see for example \cite{Speakman2001,Speakman2008}.

Much is known about bats and their evolution, yet bat evolution remains a mystery. For instance ``early evolution of the group remains poorly understood'', at least as of 2008  \cite{Wyoming}. 
 A 2001 summary of a major aspect of the puzzle  \cite{Speakman2001} reads
 ``Since there are two key behavioural traits, there are at least three scenarios for their evolution: echolocation may have evolved first ..., flight may have evolved first ..., or the two may have evolved in tandem ... This minimal view assumes that at least one of the behaviours evolved only once. Multiple origins of both traits ... could lead to much more complex interrelationships.'' The ellipsis denote reference lists. 

In this note we shall humbly suggest a resolution to this puzzle and mention possible tests.

\presub {\bf Analysis}. Let us analyze the issue of the evolution of flight and echolocation in bats and the relation between them. Together they enable flight (and hunt) in the dark. The considerable diversity of bats is a testament to the advantage offered by this ability (in fact, the speciation is known to have been going on throughout the cenozoic era, as we mention later).

If we consider each innovation separately we find that flight in its own is very useful, as clearly evidenced by the large number of flying species without echolocation including birds and insects. On the other hand, I view the rarity of terrestrial echolocating species as evidence for the low usefulness of echolocation without flight (on the ground).  Therefore I choose to continue by assuming that \emph{flight was developed first}. 

At this point I wish to pose a question: \emph{why was echolocation necessary for bats?} Before discussing the question it is perhaps worthwhile to explain what is meant by necessary.  Echolocation is clearly useful, yet it is also rather involved or specialized. Not only does it require various bat resources such as energy (for sound) and signal processing load on the brain, but also its evolution requires a large number of steps. There is evidence that all extant bats evolved from an echolocating ancestor, see e.g. \cite{Teeling2005}. Hence the considerable price of echolocation must have been outweighed by considerable advantages and selective pressures, namely by necessity. 

The problem is that echolocation appears to be \emph{unnecessary} for the following reasons. First, isn't it conceivable that the proto-bat would be active during the day (be diurnal rather than nocturnal) thereby making echolocation completely redundant? In fact, \cite{Speakman2001} has already advanced this fascinating and suggestive hypothesis. Secondly, it appears to be possible to get along at night without echolocation. In fact, nights are not completely dark, especially in the presence of moonlight, and owls are a living proof that  heightened sight and hearing can very well suffice. The same conclusion can be strengthened from a different angle: echolocation allows bats to inhabit completely dark caves. Yet, it is doubtful that caves are attractive enough to drive echolocation evolution.

In order to find a way out of this puzzle we resort to consider whether at any time in the past conditions were so markedly different from today, so that echolocation became truly necessary.

\presub {\bf A formative event for echolocating bats}. Known facts about bat evolution offer hints towards identification of such a time. The currently oldest bat fossil was dated to some 52 million years ago (Mya) during the early Eocene period \cite{Wyoming}. 
Recent genetic studies enable an improved determination of phylogenetic trees including time calibration. 
\cite{ShiRabosky2015} assembled such a phylogeny of 812 extant bat species 
  and found in particular that all extant bat species evolved from a single species around 58 Mya (though such an absolute date is subject to further calibration). Other studies are qualitatively similar, but locate the last bat ancestor at 64 Mya \cite{Teeling2005} or at roughly 90 Mya \cite{DelayedRise2007}.

These dates are in proximity to the Cretaceous--Paleogene (K--Pg) extinction which occurred some 66 Mya. During the K--Pg boundary (also known as the Cretaceous-Tertiary (K--T) extinction)  some three quarters of all plant and animal species became extinct \cite{wiki:KPg} famously destroying all non-avian dinosaurs. It is used to delineate the mesozoic (middle life) era which preceded it (252-66 Mya) from the cenozoic (new life) era, which followed it (66 Mya - present), and is known as the age of mammals. Evidence suggests that the calamity was caused by an impact of a roughly 10km rock from space into a crater currently in the Gulf of Mexico, releasing the energy equivalent to some 100 Teratonnes of TNT \cite{Alvarez1980}, see \cite{Schulte2010} for a recent review. Evidence includes high levels of Iridium in the global K--Pg clay layer.

By the way, we can now answer another question from the introduction, namely why the relative diversity of bats among mammals is so high: bats appeared rather early among mammals and they had been diversifying rather consistently ever since \cite{ShiRabosky2015}.

Following the timing hints we continue to consider the unusual conditions following the K--Pg event. A combination of direct evidence and modeling leads to the following picture \cite{Schulte2010}. Matter ejected into the atmosphere was distributed globally. Re-entering submillimeter spherules of condensed rock vapor caused an hours long intense thermal radiation pulse, turning the sky into an oven \cite{Melosh1990,ThermalShelter2004,HeatFire2013}. Evidence for a thermal pulse includes such spherules found in the K--Pg layer.  
 This pulse may have destroyed unsheltered animals and may have ignited global firestorms. Dust and soot clouds blocked sun light and through an analog of the nuclear winter effect brought darkness and cold temperatures for years to decades until the dust settled down. Additional effects include an earthquake, megatsunami and possibly acidification of the oceans.

For our purpose let us focus on darkness and reduced visibility. Already \cite{Alvarez1980} remarked ``this dust would stay in the stratosphere for several years and would be distributed worldwide. The resulting darkness would suppress photosynthesis...''   \cite{AsteroidImpacts1997} estimated that the K--Pg impact resulted in loss of vision from dust loading for months (table 1 there) and light levels below the limit of human vision. \cite{darkness} used the term \emph{darkness at noon}. 

\presub {\bf K--Pg and bats hypothesis}. Altogether, the reduced light and visibility during this period and its timing lead us to \emph{the hypothesis that echolocation was necessary to survive during the K--Pg extinction event}. By this we mean that echolocating bats and no other bats lived upon recovery from the K--Pg event and they are the ancestors of all modern bats. According to the hypothesis echolocation was developed before or even during the K--Pg boundary, but not after. If it evolved earlier echolocating bats may have lived together with non-echolocating bats (possibly diurnal) and were selected by the event. While K--Pg event models suggest that darkness lasted at most several decades and that would be too short a time for echolocation to evolve we keep an open mind to the possibility that limited visibility actually lasted considerably longer in which case echolocation may have evolved through accelerated evolution in the face of strong selective pressures.

The hypothesis explains why bats are echolocating (more precisely, all modern bats are descendants of echolocating bats), even though echolocation had evolved separately from flight and after it. It also explains their nocturnality which in a sense is derived from the advantages of echolocation. In the sense of the Introduction section it gives a reason for bats.  However, it does not relate to the origin of flight in bats. This hypothesis is the central conclusion of this paper. If true, then bats are a living reminder of the K--Pg event.

The hypothesis is in some tension with the conclusion of \cite{Wyoming} that the 52 Myr fossil lacked echolocation. However, this conclusion was disputed \cite{WyomingDisputed}, so we suggest to wait for additional data.

There are additional reasons to believe that the K--Pg event was key for bat evolution. With reference to Noah's ark, we may refer to the species that were selected by the event and survived as having been aboard the \emph{K--Pg ark}. The extent of the extinction is a measure of the high selection to enter this ark. In addition to echolocation it appears reasonable that flight, an insect diet and small body size were helpful for bats to survive. Indeed, it is known that of all dinosaurs only some flying (avian) dinosaurs survived (later to evolve into modern birds). For instance, flight and small body size would have been useful to find shelter in caves or burrows which according to the thermal shelter hypothesis \cite{ThermalShelter2004} was necessary to survive the thermal pulse. Flight could have been helpful also to survive the firestorms, if these occurred. Secondly, it appears likely that insects fared considerably better than vertebrate owing to their higher resilience. That would reward an insect diet and echolocation hunting. 

\presub {\bf Tests}. Having arrived at our hypothesis let us discuss several tests for it. These tests fall into two categories: bat evolution, and post K--Pg conditions.

The hypothesized bat evolution could be tested by discoveries of fossils or through decisive time-calibrated phylogenetic trees. As already mentioned, the earliest known bat fossils are dated to some 52 Mya during the early Eocene period (which occurred after the K--Pg extinction). However, due to the incomplete nature of the fossil record this only sets a bound for their appearance, and is thus fully consistent with the conjecture. In fact, this principle is known in paleontology as the Signor-Lipps effect (sometimes referred to as the inverse SL effect) \cite{wiki:Signor-Lipps}.

At the same time the hypothesis has implications for the post-K--Pg conditions including how limited visibility was, over how much of Earth and for how long.

\presub {\bf Discussion}. The hypothesis may sound far fetched. However, we know that when you have eliminated the impossible, whatever remains, however improbable, must be the truth \cite{Holmes}.

Let us discuss how the hypothesis might fail. It is possible that echolocation evolved under less extreme conditions unrelated to an extended period of global reduced visibility. If that were to be the case, then presumably one would have to look elsewhere to explain what made all bats echolocating and nocturnal. Secondly, the hypothesis puts the duration of darkness is some tension -- on the one hand it must have been long enough to affect bat evolution, yet on the other hand it must have have been moderate enough to sustain some life aboard the K--Pg ark.

It would be interesting to complete the picture regarding bats with information regarding the evolution of nocturnal birds, namely the owls, who share their habitat of the night sky. According to \cite{wiki:owls} owls were already present as a distinct lineage some 60--57 Mya, hence, possibly also some 5 million years earlier, at the K--Pg event, which makes them one of the oldest known groups of non-Galloanserae landbirds (Gallanserae are fowl). This raises the possibility that owls were promoted by the same conditions which promoted bats.

Finally, this note is written by a physicist, non-professional in evolutionary biology, and is intended to a mixed and general scientific audience. It is rather likely that experts could detect errors small or large, and admittedly it is a hypothesis and not a proven result. With that in mind, I, your humble patience pray, Gently to hear, kindly to judge, our play \cite{Henry}. 

\subsection*{Acknowledgments}

It is a pleasure to thank Sharon Swartz, Doug Robertson and John Speakman for very useful comments. 

\end{document}